\documentclass{PoS}
\pdfoutput=1

\usepackage{graphicx}

\newcommand{\bc}{\begin{center}}
\newcommand{\ec}{\end{center}}

\newcommand{\bd}{\begin{description}}
\newcommand{\ed}{\end{description}}

\newcommand{\bi}{\begin{itemize}}
\newcommand{\ei}{\end{itemize}}

\newcommand{\be}{\begin{enumerate}}
\newcommand{\ee}{\end{enumerate}}

\newcommand{\bq}{\begin{quote}}
\newcommand{\eq}{\end{quote}}

\def\usec{\mu\mbox{\hspace*{-.1mm}sec}}

\title{An FPGA-based Torus Communication Network}\ShortTitle{An FPGA-based Torus Network}

\author{\vspace*{-5mm} \hfill {\tt DESY 11-011}\\~}

\author{\speaker{Marcello Pivanti}     \\ 
        INFN and University of Ferrara,
        Via Saragat 1, I-44100 Ferrara, Italy\\
        \email{pivanti@fe.infn.it}}

\author{Sebastiano Fabio Schifano      \\
        INFN and University of Ferrara,
        Via Saragat 1, I-44100 Ferrara, Italy\\
      	\email{schifano@fe.infn.it}}

\author{Hubert Simma   \\
      	NIC, DESY, Platanenallee 6, D-15738 Zeuthen, Germany  \\
      	\email{hubert.simma@desy.de}}

\abstract{
We describe the design and FPGA implementation 
of a 3D torus network (TNW) to provide nearest-neighbor communications 
between commodity multi-core processors. The aim 
of this project is to build up tightly interconnected and scalable 
parallel systems for scientific computing. The design includes the 
VHDL code to implement on latest FPGA devices a network processor, 
which can be accessed by the CPU through a PCIe interface and
which controls the external PHYs of the physical links. Moreover,
a Linux driver and a library implementing custom communication APIs 
are provided.
The TNW  has been successfully integrated in two recent parallel 
machine projects, QPACE and AuroraScience. We describe some 
details of the porting of the TNW for the AuroraScience system
and report performance results.

}

\FullConference{
The XXVIII International Symposium on Lattice Field Theory, 
Lattice2010       \\
June 14-19, 2010  \\
Villasimius, Italy
}

\begin{document}


\section{Introduction}

One of the key elements of a massively parallel computer 
is the communication network that interconnects the computing nodes. 
It allows the 
processes  running on the CPU cores to cooperate as a large unique 
entity to solve computational problems fast. 
Parallel scientific computing often requires machines with tightly-coupled nodes, 
allowing applications, like Lattice-QCD or Lattice-Boltzmann, to efficiently perform 
fine-grained communications and to scale in the {\em strong regime}, i.e. at constant 
problem size.
Several massively parallel machines optimized for Lattice-QCD simulations
have been developed by user-community projects, such as for example
the APE systems \cite{apecise} in Europe, and the machines of the 
Columbia University~\cite{columbia} in the US.
Both developments have been based on a custom design of processor and network.
To efficiently support the most relevant communication patterns of Lattice-QCD 
algorithms, a common choice for the network topology is a $D$-dimensional mesh 
($D=3,4,\ldots$) with periodic boundaries.
The pairwise connectivity between (nearest-neighbour) nodes avoids the typical
bottlenecks of switched networks and allows to scale machine performance 
to thousands of processes (until global reductions may become relevant).

Modern commodity processor architectures, like x86 with SSE or Cell BE, can be 
efficiently exploited for LQCD computations~\cite{luscher,bilardi}
and have low cost and power consumption per flop.
The option of using off-the-shelf CPUs has lead to a new strategy
in designing LQCD-optimized parallel machines, based on standard
multi- or many-core processors interconnected by a {\em custom} network. 
This approach has been exploited in the QPACE~\cite{qpacecise08,qpacepower} 
and AuroraScience~\cite{paperoLuigi} projects.

The torus network (TNW) which we have developed for these machines 
provides a simple but efficient interconnection network between multi-core commodity processors 
and can easily be ported to other architectures which support a 
standard IO technology, like PCI-express (PCIe), to interconnect CPU 
and network processor.
Data transmission is based on a light-weight custom protocol with 
minimal overheads from operating system or software layers.
The logics of the network processor is tailored to be implemented on 
recent {\em Field Programmable Gate Array} (FPGA) devices.
This allows a quick and flexible development avoiding the risks 
and non-recurrent costs of an ASIC implementation.

The hardware building block of both machines, QPACE and AuroraScience, 
is a compact node card which, apart from basic peripheral components, 
hosts the CPU(s), the RAM, and an FPGA with 6 external transceiver 
devices (PHY) to drive the 6 TNW links, each with a bit rate of 10\,Gbit/s.
In QPACE, each node is equipped with one IBM Cell processor (PowerXCell 8i) 
connected through the FlexIO bus with the FPGA  (Xilinx Virtex-5 LX110T). 
The FPGA acts as south-bridge, Ethernet controller, and network processor for 
the torus network. The QPACE machines are integrated with a novel 
liquid cooling system and ranked as top entry of the GREEN 500 list 
in November 2009 and June 2010. 
The nodes of the AuroraScience machine are based on latest Intel 
multi-core CPUs. Each node has two six-core CPUs 
(four cores in the first version), 12\,GByte of RAM. and a south-bridge. 
It is connected to the FPGA (Altera Stratix IV GX-230)
by two Gen2 8x PCIe interfaces, each providing 
an effective bandwidth of 3.2\,GByte/s. In addition to the torus network, 
the nodes of AuroraScience are interconnected by a switched InfiniBand 
network.

In the following we explain the concepts and architecture of the TNW,
and discuss some implementation details of the TNW as used in AuroraScience. 
We briefly describe the system software for the TNW (driver and communication 
library) and report on early performance results.


\section{TNW Architecture}
%
%
The FPGA of each node implements a network processor (NWP) which
is the interface between the CPUs (or south-bridge) of a node and 
the TNW links. The NWP provides the hardware control of the data
transmission and has {\em injection} and {\em reception} buffers 
for each of the 6 links. 

%
%
Applications access the TNW by (i) moving data into the injection
buffer of the NWP of the sending node, and (ii) enabling data to
be moved out of the reception buffer of the NWP of the receiving node.
Thus, the data transfer between two nodes proceeds 
according to a {\em two-sided} communication model, i.e. explicit 
operations of both CPUs, sender and receiver, are required to control the data 
transmission of a message. 

Tracing a CPU-to-CPU data transfer over the TNW the following 
three steps occur:
\\
1. The send operation simply moves the data items of a message into the 
injection buffer for one of the 6 links in the NWP of the sending node. 
Depending on the architecture
and IO-interface of the CPU, this operation can be implemented
according to different schemes (see sect. ~\ref{procint}).
\\
2. As soon as an injection buffer holds data, the NWP breaks it into 
fixed-size packets and transfers them in a strictly ordered and 
reliable way over the corresponding link. Of course, the transfer 
is stalled when the reception buffer of the destination NWP runs 
out of space (back-pressure). 
\\
3. The receive operation on the destination CPU is initiated by passing 
a {\em credit} to its NWP. The credit provides all necessary control
information to the receiving NWP to move the received data packets
to the destination CPU and to {\em notify} it
when the last packet of a message has been delivered.

%
%
To allow a tight interconnection of processors with a 
multi-core architecture, the TNW also supports the concept of 
{\em virtual channels} to multiplex multiple data streams over 
the same physical link. 
A virtual channel is identified by an index (or tag) which is transfered 
over the link together with each data packet.
This is needed to support independent message streams between different pairs
of sender and receiver threads (or cores) over the same link. 
The virtual channels can also 
be used as a tag to distinguish independent 
messages between the same pair of sender and receiver threads.
Currently the TNW design supports $8$ virtual channels, but 
this number can be increased, e.g. to support CPUs with more cores,
at the expense of additional resource usage on the FPGA.

%
%
The simple communication model of the TNW requires that
each send operation has a corresponding receive operation. Moreover,
send operations which refer to the same link and virtual channel must 
be issued in the same order as the corresponding receive operations.

\begin{figure}[t]
\center
\begin{minipage}[c]{.62\textwidth}
\includegraphics[width=\textwidth]{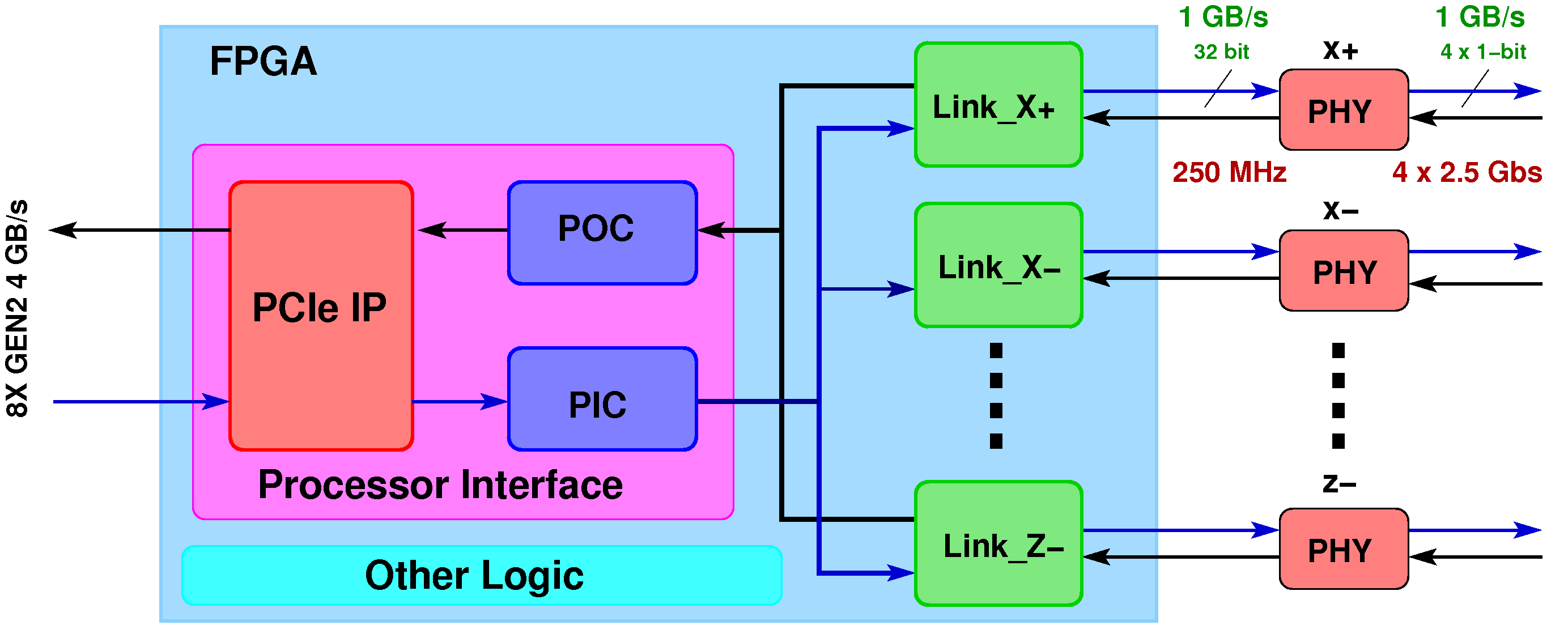}
\end{minipage}
\hfill
\begin{minipage}[c]{.34\textwidth}
\includegraphics[width=\textwidth]{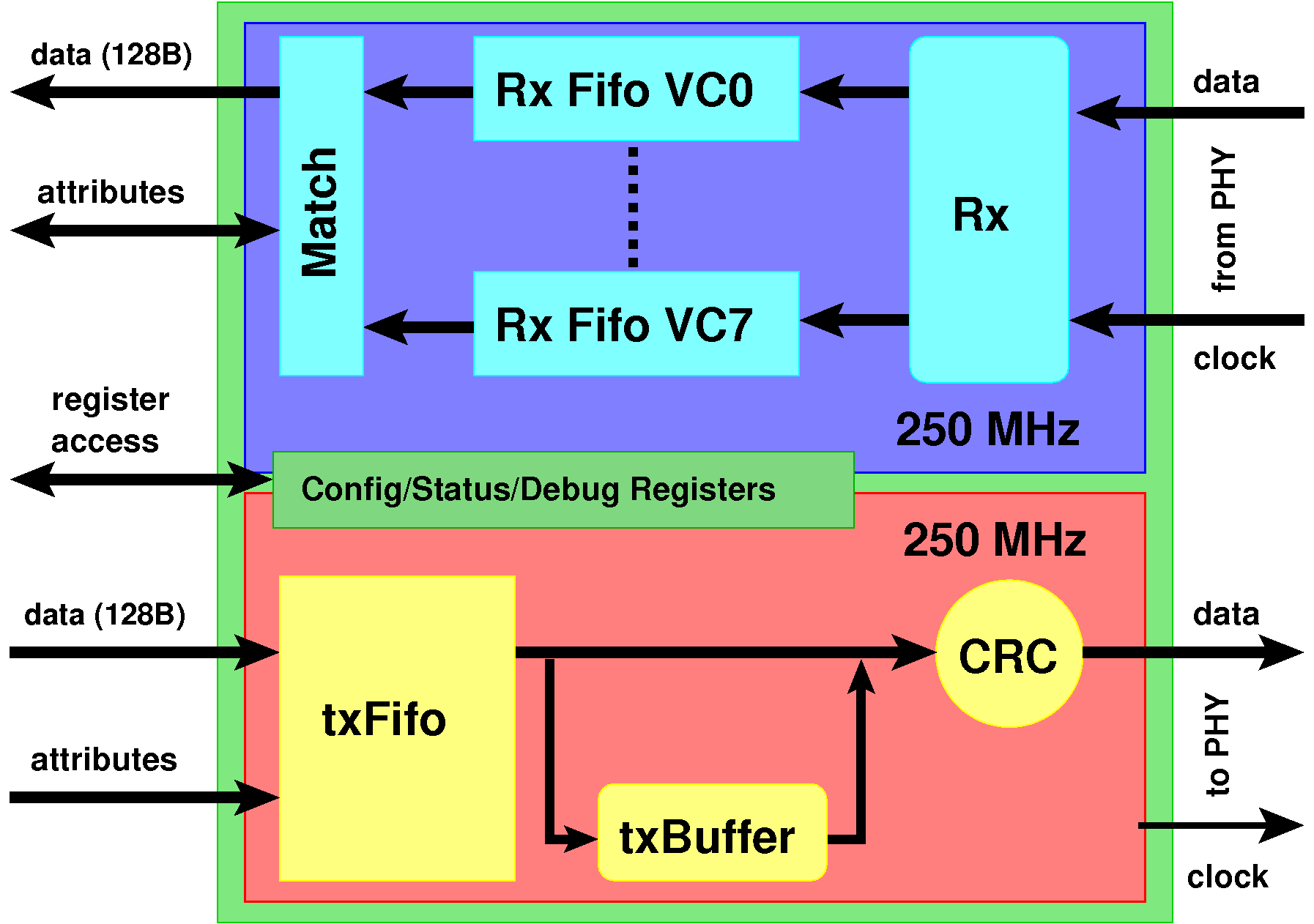}
\end{minipage}
\caption{Block Diagrams of the network processor (left) and link module (right)}
\label{nwp-block-diagram}
\end{figure}

The architecture of the NWP is shown in figure~\ref{nwp-block-diagram} (left part). 
The main logic blocks, which will be described in more detail in the following, are 
the processor interface and 6 link modules.

\subsection{Processor Interface}\label{procint}
To move data, as well as control or status information 
between NWP and CPU, the processor interface handles
inbound transactions (initiated by the CPU) 
and generates outbound transactions (initiated by the NWP)
according to the specific IO protocol supported by the CPU, 
e.g. PCIe. 

Most of the basic data movements correspond in an 
obvious way to either inbound or outbound transactions.
For instance, writing or reading registers of the NWP
typically corresponds to inbound transactions. The
most natural and efficient way to implement the (non-blocking)
receive operation is by moving the received 
data by outbound transactions from the NWP to the CPU 
(followed by a final outbound transaction to notify the 
CPU when the receive operation is completed).
However, to implement the send operation two different
schemes, called {\sf Pput} and {\sf Nget} in the following, 
can be convenient for moving the data from the CPU to the NWP. 

In the {\sf Pput} scheme, the CPU initiates and controls
the data transfer to the NWP. It is then convenient
to map the injection buffers of the different links and
virtual channels directly into disjoint areas of the address 
space of the CPU. Then, a single IO-transaction can be sufficient 
to move to the NWP both, the data and all control information 
(which can be implicitly encoded into the addresses). 

On the other hand, in the {\sf Nget} scheme the data transfer is
controlled by the NWP. Therefore, the CPU first has to pass the 
required control information to the NWP and finally the NWP has 
to notify the CPU that the data transfer is completed (i.e. the 
application can re-use the memory locations from where the sent 
data originated).

Compared to implementing the send operation through an {\sf Nget} scheme, 
a {\sf Pput} scheme can be more efficient, 
in particular for short messages, because it requires fewer
IO-transactions and hence may have a lower latency. 
However, in the {\sf Nget} scheme it can be simpler to handle {\em back-pressure}, 
which arises when injection buffers are full, while this may require extra
transactions in the {\sf Pput} scheme.
In QPACE non-blocking send operations are implemented according to the {\sf Pput} scheme 
exploiting the ability of the SPE cores to perform {\em Direct Memory Access} (DMA) operations.

The TNW implementation on the Altera Stratix IV FPGA, as used in AuroraScience,
has a PCIe-based processor interface. It consists of the
three major blocks indicated in figure~\ref{nwp-block-diagram}.
The {\sf PCIe IP} is a hardware macro embedded in the 
FPGA and provides the PCIe protocol stack (physical, 
data link, and transaction layers). 
Two application-logic blocks, {\sf PIC} and {\sf POC}, are attached
to the {\em Avalon} interface of the IP macro. They handle 
incoming (down-stream) transaction layer packets (TLP) 
and generate outgoing (up-stream) TLPs, respectively.

In contrast to the Cell processor, where each SPE core has a DMA engine, 
on x86 CPU architectures the {\sf Pput} scheme needs to be implemented 
by {\em Programmed IO} (PIO), i.e. the CPU performs store operations 
to the memory addresses, where the injection buffers have been mapped.
The memory system and south bridge translate these store operations 
to individual {\em memory-write} PCIe transactions with a small
payload of only 16 Bytes (the size of an SSE register).

To avoid these inefficient partial writes and improve performance, one 
can enable {\em write-combining} (WC) memory transfers. Stored data is then written 
into small temporary buffers of the CPU. When such a WC buffer is full, 
it is flushed to the IO-bus by a single burst 
transfer with a payload of 64 Bytes (the size of a cache-line).
Since the WC buffers may be flushed also for other reasons, like thread de-scheduling,
message fragments can arrive at the NWP in an interleaved order. Therefore, 
the NWP must implement a re-order logic which restores the correct order
of data before it is moved into the injection-buffer.

On x86, with a PIO-based {\sf Pput} scheme the send operations 
can only be implemented in a blocking way.
Therefore, we are developing extensions
of the NWP to also support the {\sf Nget} scheme. This
allows to exploit DMA mode 
and to have non-blocking send operations.

\subsection{Link Module}
The link modules, one for each of the six directions of the 3D torus, 
control the transfer of data packets over the physical links of the TNW. 
Each link module on the FPGA is connected through a 32-bit bus to an 
external transceiver PHY (PMC Sierra PM8358a).
The PHYs of neighbor nodes are connected by high-speed serial links (XAUI) 
which are routed over backplane and/or cable. 
The link module implements a light-weight and robust custom protocol 
to guarantee data integrity and strict ordering of the packets
by making use of the control symbols from 8/10b coding by the PHY. 
Each TNW link can simultaneously send and receive data with
a maximal data throughput of 0.91\,GByte/s per direction 
(taking into account the overhead from coding and custom protocol).

The architecture of a link module is shown in the right part of figure~\ref{nwp-block-diagram}
and consists of two basically independent parts for sending and receiving.
The data to be sent is stored together with all relevant control 
information, like address offset and virtual channel index, in
the injection buffer ({\sf txFifo}), which is handled with First-In-First-Out
policy.
As soon as it holds at least 128 Bytes of data,  
the transmission logic starts to pop data from {\sf txFifo} and generates a packet
composed of a 32-bit header, 128 Bytes of data payload (32$\times$32 bit), 
and a 32-bit CRC. The packet is then passed to the external PHY for 
transmission over the physical link.

On the other side, the {\sf Rx} logic decodes the data packets received 
from the external PHY, re-computes the CRC, and compares it with the 
CRC of the packet.
If they match, the packet is pushed into the reception FIFO associated 
with each virtual channel, and a positive feedback ({\sf ACK}) is sent 
back over the link to the link module of the sender. If the CRC does 
not match, or in case of other errors in receiving the packet, a 
negative feedback ({\sf NACK}) is sent back, and all further 
data packets from the PHY are discarded until a {\sf RESTART} command 
is received.

The link module temporarily stores each packet, which has been sent, 
into an internal buffer ({\sf txBuffer}) until a feedback for that 
packet is received from the peer-link. 
If the feedback is positive the corresponding packet is dropped, 
otherwise the link module recursively enters in {\em resend} mode, 
sends a {\sf RESTART} to the peer, and then begins to 
re-send all packets from {\sf txBuffer} until a positive feedback 
has been received for all of them and further data from the 
injection buffer can be handled.

The reception buffer is logically organized as separate \mbox{FIFOs}
for each virtual channel. As soon as a credit provided by the destination 
CPU matches a 128-Byte data packet stored in the reception buffer, 
the receiver logic signals to the processor interface that data is ready
and provides the memory address where to store it. The 
processor interface then transfers the data from the reception buffer
into the memory of the CPU. Multiple credit--packet matches are arbitrated 
according to a round-robin policy.      


\section{System Software}

To provide convenient and efficient access to the TNW from thread-applications, 
we have developed a Linux driver and a basic communication library.

For the efficient support of the {\sf Pput} scheme, the driver marks 
the address space of the injection buffers (allocated at boot 
time by the NWP) as write-combining and maps them into the address space
of the threads by standard {\em mmap} kernel functions. 
Thus, threads can directly access the injection buffers avoiding time 
overhead from frequent context switches between user- and kernel-mode.
For receiving data and notifications from the NWP, the driver allocates 
contiguous memory areas for each virtual channel.
Control and status registers of the NWP are mapped  
on separate addresses
and are accessible directly from user-space.

The communication library implements a custom API and provides 
functions to send and receive messages, and to configure, control, and monitor 
the behavior of the NWP. The most relevant communication functions
are {\sf tnwSend} to send a message, {\sf tnwCredit} to initiate 
a receive operation, as well as functions to test or wait (poll) for the 
completion of receive (and send) operations.

The structure of many scientific codes is {\em Single Program Multiple Data} (SPMD). 
Typical communication patterns, like sending on all nodes data in direction $x$+
and receiving from $x$-, are readily translated to our API: {\sf tnwCredit} to initiate 
receiving from $x$- and {\sf tnwSend} to start transmission to $x$+, 
followed by tests for completion of the transfer and availability of the data.


\section{Results and Conclusions}

\begin{figure}[t]
\center
\includegraphics[scale=0.33]{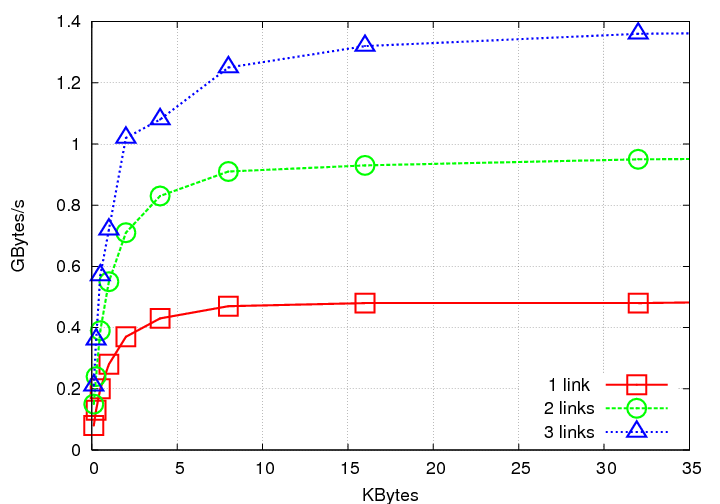}
\caption{Aggregate bandwidth of 1, 2, and 3 links in the TNW implementation on AuroraScience.}
\label{aggregatebw}
\end{figure}

The TNW is implemented as a highly pipelined design which runs on the 
Altera Stratix IV GX-230 at a frequency of 250 MHz, using about 18\% 
of the logic and 17\% of the embedded memory.
Taking into account protocol overhead, each TNW link has a 
maximal theoretical bandwidth of 0.91\,GByte/s. 
Figure~\ref{aggregatebw} shows the effective bandwidth of the 
TNW which scales well when data is simultaneously sent 
over 1, 2, and 3 links.
This implementation uses the {\sf Pput} scheme
and the bandwidth does not exceed $\approx 0.5$ GB/s per link
when flushing of WC buffers is software controlled.

On the other hand, when flushing of WC buffers is handled by the re-order logic 
in the NWP, the {\sf Pput} scheme 
achieves a maximal transmission bandwidth of 0.83\,GBbyte/s per link, 
i.e. 90\% of the theoretical bandwidth, and has an average 
latency of $1.67\,\usec$ (measured as transfer time of a message with 
128 B payload).
In a first test implementation of the Nget scheme, we find a maximal 
bandwidth of 0.76\,GByte/s and an average latency of $2.16\,\usec$.
The NWP-to-NPW latency, i.e. the time from the instant when a packet enters the NWP 
to the instant when it is ready to be delivered to the CPU, is about $0.6\,\usec$ 
(of which $0.24\,\usec$ arise from the PHYs).
Since each PCIe transfer to or from the CPU memory causes about $0.2\,\usec$ additional latency,
we estimate a software overhead of $\approx 0.7\,\usec$ for the 
{\sf Pput} scheme, and $\approx 0.8\,\usec$ for the {\sf Nget} scheme.

As an application test of the TNW and of the communication primitives we have adapted 
a 2D fluid-dynamics simu\-lation code based on the Lattice-Boltzmann method.
This code has also been used on QPACE~\cite{iccs10}, and we have
optimized it for AuroraScience. On a 16-node system we obtain
about 36--39\,\% of the peak performance of 160\,Gflops double-precision per node.

In this paper we have presented the design of a torus network
which provides efficient nearest-neighbor communications and
allows fine grained parallel programming.
The TNW design has been successfully integrated into the QPACE 
and AuroraScience machines, and we plan to make it available 
as open source project \cite{ftnw}.
An optimized implementation of the 
{\sf Nget} scheme is under development \cite{nget}. 
Future work may explore improved link technologies or functional 
extensions, like support for more general communication patterns.

\vspace*{5mm}

{\bf Acknowledgements: }
We like to thank the members of the QPACE project, in particular 
T.~Mauer, A.~Nobile, D.~Pleiter, and T.~Streuer for important 
contributions to the design of the TNW and for their strenuous 
efforts to integrate and test it in QPACE.
We thank K.H.~Sulanke for the design of the 
test-boards on which most of the TNW logic has been developed. 
We are grateful to R. Tripiccione for many discussions and advices
from the APE experience. We thank AuroraScience for stimulating 
our work.
M.P. has been supported by DFG (SFB/TR55) and H.S. by University Milano Bicocca.


\end{document}